\documentclass[aps,prl,twocolumn,showpacs,superscriptaddress,groupedaddress]{revtex4}
\usepackage{graphicx}
\usepackage{dcolumn}
\usepackage{bm}
\usepackage{amssymb}
\usepackage{subfigure}
\usepackage{sidecap}
\begin{document}
\title{Hyperentanglement-enabled Direct Characterization of Quantum Dynamics}
\author{T.M.~Graham} \affiliation{Department of Physics, University of Illinois at Urbana-Champaign,Urbana, Illinois 61801-3080, USA}
\author{J.T.~Barreiro} \affiliation{Institut f\"ur Experimentalphysik, Universit\"at Innsbruck, Technikerstrasse 25, A-6020 Innsbruck, Austria}
\author{M.~Mohseni} \affiliation{Research Laboratory of Electronics, Massachusetts Institute of Technology, Cambridge, Massachusetts 02139, USA}
\author{P.G.~Kwiat} \affiliation{Department of Physics, University of Illinois at Urbana-Champaign,Urbana, Illinois 61801-3080, USA}

\begin{abstract}
We use hyperentangled photons to experimentally implement an entanglement-assisted quantum process tomography technique known as Direct Characterization of Quantum Dynamics. Specifically, hyperentanglement-assisted Bell-state analysis enabled us to characterize a variety of single-qubit quantum processes using far fewer experimental configurations than are required by Standard Quantum Process Tomography (SQPT). Furthermore, we demonstrate how known errors in Bell-state measurement may be compensated for in the data analysis. Using these techniques, we have obtained single-qubit process fidelities as high as 98.2\% but with one-third the number experimental configurations required for SQPT. Extensions of these techniques to multi-qubit quantum processes are discussed.
\end{abstract}

\maketitle

As scientists advance the frontiers of quantum information science and quantum computing by producing ever larger and more complex quantum systems, there has been an increased need for efficient methods of characterizing quantum states and processes.  The information contained in a quantum system may be extracted by a technique known as Quantum State Tomography (QST), which is accomplished by making various measurements on multiple copies of the state, and then using these measurement outcomes to reconstruct the density matrix.  Similarly, the information describing a quantum process is extracted by probing the process with various quantum states and then making measurements on the output.  This characterization, known as Quantum Process Tomography (QPT), is generally more difficult than QST because quantum processes contain quadratically more information than the states on which they operate.  Because of this difficulty, many different techniques have been invented to characterize quantum processes.  We present an experimental realization of a QPT technique devised by Mohseni and Lidar \cite{DCQD}, known as Direct Characterization of Quantum Dynamics (DCQD). DCQD has the advantage over other QPT methods in that it both requires far fewer experimental settings than techniques which use only local probe states and measurements, and it requires less complicated---but physically realizable---measurements than other techniques requiring a similar number of experimental configurations.

The biggest challenge in applying DCQD for optical qubits is performing the required full Bell-state analysis (BSA) on each output state, which is impossible using only linear optics and a restricted Hilbert space \cite{BSA}. DCQD was implemented with photons by Wang et al. using a probabilistic BSA \cite{probDCQD}; however, the lack of full BSA meant that substantially more measurements per experimental configuration were required, losing much of the DCQD advantage. However, we have previously shown that it is possible to achieve full BSA using quantum systems that are “hyperentangled”---simultaneously entangled in multiple degrees of freedom \cite{hyperentanglement}.  In fact, DCQD using such deterministic BSA was demonstrated by Liu et al. \cite{hybridDCQD}, but was achieved using only single-photon \textit{hybrid}-entangled states (entanglement between different degrees of freedom of a single particle), and the techniques used could not be scaled to the multi-photon case. 

Here we demonstrate DCQD using hyperentangled-enabled deterministic BSA and techniques that can readily be extended to characterize higher dimensional quantum processes. Specifically, we used photons simultaneously entangled in both polarization and orbital angular momentum to characterize several classes of single-qubit polarization quantum processes using DCQD.  By using hyperentangled photons to implement DCQD, we were able to characterize single-qubit quantum processes with only one-third the number of experimental configurations than would be required using SQPT.  After discussing the Standard Quantum Process Tomography (SQPT) methods for comparison, we will describe the essential elements of DCQD. Next, we describe our experimental implementation of both SQPT and DCQD, before discussing some possible extensions of the latter techniques.

The information describing a quantum process may be completely parameterized by the $\chi$-matrix, a super-operator that maps an input quantum state to an output quantum state; for a single qubit \cite{SQPT}:
\begin{equation}
\epsilon(\rho)=\sum\limits_{m,n=0}^{3}\chi_{mn}\sigma_{m}\rho\sigma_{n},
\end{equation}
where a quantum process, $\epsilon$, acting on a quantum state, $\rho$, can be represented as sum of transformations made by Pauli matrices, $\sigma_m$, weighted by the elements $\chi$ \cite{SQPT}.  With SQPT, the technique commonly used to characterize quantum processes, a QST is performed on a complete set of input states (Fig. 1(a)).  SQPT has the advantage of requiring only simple input states and simple measurements. However, this technique requires $12^n$ experimental configurations to completely characterize a quantum process acting on $n$ qubits [($4^n$ input states)$\times$($3^n$ measurement settings)], which rapidly becomes intractable as n increases \footnote{It is assumed that all outcomes of projective measurements may be detected (i.e. H\&V (and D\&A and L\&R) can both be simultaneously measured)}.

\begin{figure}[]
\label{sqpt}
\subfigure{\includegraphics[width=\linewidth]{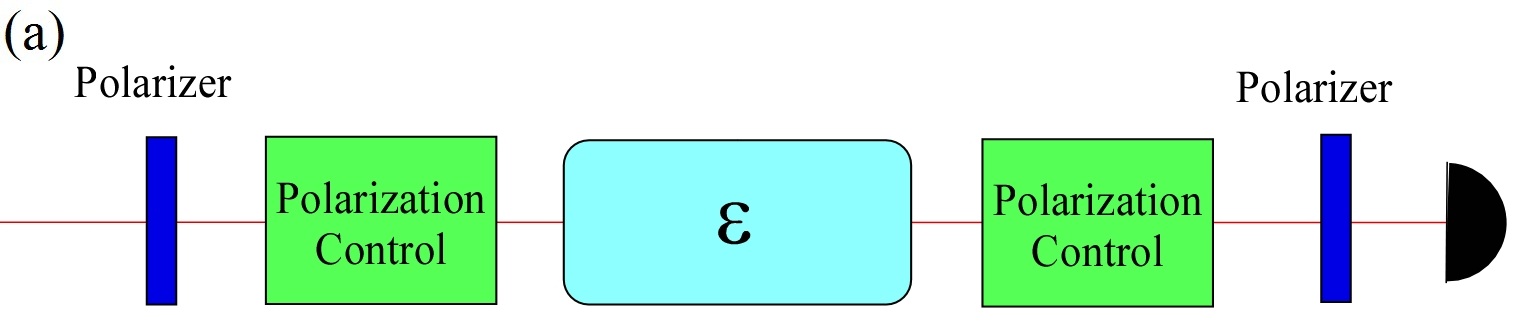}}\\
\subfigure{\includegraphics[width=\linewidth]{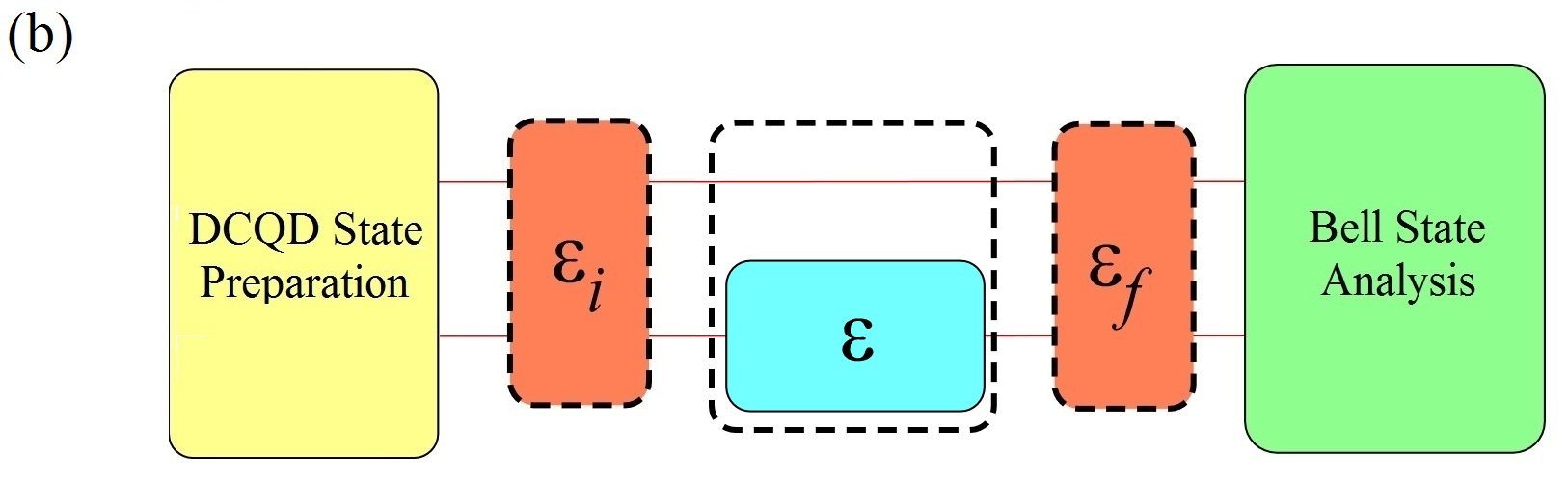}}
\caption{
The basic measurement schemes for SQPT (a) and DCQD for a single-qubit process $\epsilon$ (b). Though DCQD requires more complex input states and measurement settings than SQPT, it also requires four times fewer experimental configurations. The dotted boxes before and after the process in (b) represent errors in source preparation and analysis, here described by initial and final error processes ($\epsilon_i$ and $\epsilon_f$).}
\end{figure}

Another type of QPT---Ancilla Assisted Process Tomography (AAPT)---uses the nonlocal behavior of entangled qubits to decrease the number of necessary inputs to one. Here, the input qubits are entangled with an equal number of ancilla qubits \cite{AAPT}; by performing QST on the total quantum state, it is possible to completely reconstruct the quantum process.  In fact, by making Mutually Unbiased Basis (MUB) measurements or using Positive Operator Value Measures (POVM) instead of simple Joint State Measurements (JSM), it is possible to use AAPT to perform the QST and reconstruct the quantum process with far fewer experimental settings than are required by SQPT or even DCQD \cite{resource}. However, the complexity of the measurements required to perform these process tomography techniques increases as the number of qubits $n$ increases, because they require many-body interactions (which are difficult with any current qubit technology, and are \textit{impossible} to implement deterministically with linear optics).

Unlike SQPT and AAPT, DCQD can \textit{directly} characterize a quantum process without QST; the process is characterized by performing a full BSA on a specific set of partially entangled quantum states whose ancilla qubits have interacted with the quantum process (Fig. 1(b)).  In fact, a judicious choice of states also allows one to minimize errors in the process estimation \cite{GDCQD}. Not only does the number of experimental settings required by this technique ($4^n$) scale better than SQPT ($12^n$), but the most complicated measurement required is a 2-qubit BSA, no matter how complicated the process to be measured (i.e., even a 4-qubit process requires only pairwise BSA).

We used SQPT and DCQD to measure a variety of single-qubit quantum processes which act on photon polarization.  For both of these techniques, the quantum single- and two-photon polarization states necessary for each respective technique were produced using Type I spontaneous parametric down-conversion.  Specifically, time-correlated 702-nm photon pairs were created by pumping a pair of $\beta$-barium borate (BBO) crystals with 351-nm light from an Ar$^+$ laser.  In the SQPT measurements, the idler photon was used to herald the presence of a horizontally polarized single signal photon.  Liquid crystals were used to prepare the single photons into one of an over-complete set of probe states \cite{remote}: (horizontal (H), vertical (V), diagonal (D), anti-diagonal (A), left circular (L), and right circular (R)).  After transmitting each input state through the process, an over-complete set of measurements was performed on each resulting output state (in the same bases as above) using an adjustable quarter- and half-wave plates before a polarizer.  Though it is possible to characterize a single-qubit quantum process using only 12 experimental configurations, we used a full 18 experimental configurations to minimize the error in the SQPT data, since it was used as a reference for the DCQD measurements \cite{36vs16, 36vs16p2}.

Measuring quantum processes using DCQD requires entangled input states and measurements.  The same downconversion source was used for this technique, except now both crystals were pumped coherently with a superposition of H and V polarizations, which created polarization entangled photon pairs \cite{SPDC} (only partially for some input polarizations \cite{nonmax}). In addition, due to conservation of orbital angular momentum, the photons were entangled in spatial mode \cite{spatial}.  Pumping the crystals with equal parts of H and V polarization prepares the following hyperentangled state \footnote{In these experiments we discard all other spatial modes, e.g., both photon Gaussians.}:
\begin{equation}
\frac{1}{2} ( | HH \rangle - | VV \rangle ) \otimes ( | \circlearrowleft \circlearrowright \rangle + | \circlearrowright \circlearrowleft \rangle ),
\end{equation}
where $\circlearrowleft$ and $\circlearrowright$ represent right- and left-orbital angular momentum modes (OAM), respectively. Using a pump beam polarized at $\frac{\pi}{8}$ with respect to H and by manipulating the polarization of both signal and idler photons using liquid crystals, it was possible to create the three additional input states required for DCQD:
\begin{eqnarray}
&&\frac{1}{\sqrt{2}} (\cos\frac{\pi}{8}|HH\rangle-i\sin\frac{\pi}{8}|VV\rangle)\otimes(|\circlearrowleft\circlearrowright\rangle+|\circlearrowright\circlearrowleft\rangle)\nonumber\\
&&\frac{1}{\sqrt{2}} ( \cos \frac{\pi}{8} | DD \rangle -i \sin \frac{\pi}{8} | AA \rangle ) \otimes ( | \circlearrowleft \circlearrowright \rangle + | \circlearrowright \circlearrowleft \rangle )\nonumber\\
&&\frac{1}{\sqrt{2}} ( \cos \frac{\pi}{8} | LL \rangle -i \sin \frac{\pi}{8}|RR \rangle ) \otimes ( | \circlearrowleft \circlearrowright \rangle + | \circlearrowright \circlearrowleft \rangle ).
\end{eqnarray}
Only the signal photon of each state was then propagated through the quantum process, after which we used hyperentanglement to allow us to perform a full polarization BSA on the output states \cite{hyperBSA} (Fig. 2(a)). 

The key point to full BSA is the fact that a hyperentangled polarization spatial-mode Bell state may be written \cite{hyperBSA}:
\begin{eqnarray}
\Phi^\pm_{Spin} \otimes \Psi^+_{Orbit} &=& \frac{1}{2}(\phi^+_1\otimes\psi^\pm_2 + \phi^-_1\otimes\psi^\mp_2\nonumber
\\ && + \psi^+_1\otimes\phi^\pm_2 + \psi^-_1\otimes\phi^\mp_2)\nonumber \\
\displaystyle \Psi^\pm_{Spin} \otimes \Psi^+_{Orbit} &=& \frac{1}{2}(\pm\phi^+_1\otimes\phi^\pm_2 \mp \phi^-_1\otimes\phi^\mp_2\nonumber
\\ && \pm \psi^+_1\otimes\psi^\pm_2 \mp \psi^-_1\otimes\psi^\mp_2),
\end{eqnarray}
where $\Phi^\pm$ and $\Psi^\pm$ represent the four two-photon Bell states for polarization (spin) and orbital angular momentum (orbit), and $\phi^\pm$ and $\psi^\pm$ are the single-photon hybrid-Bell states for both signal (1) and idler (2) photons:
\begin{eqnarray}
&&\psi^\pm=( | H\circlearrowleft \rangle \pm | V\circlearrowright \rangle )/\sqrt{2}\nonumber\\
&&\phi^\pm=( | H\circlearrowright \rangle \pm | V\circlearrowleft \rangle )/\sqrt{2}.
\end{eqnarray}
Thus, determining the single-photon hybrid-entangled state of each individual photon uniquely determines the two-photon hyperentangled state.  The requisite operations were implemented by interfering the $\pm 1$-order diffracted outputs of a forked hologram on a polarizing beam splitter (Fig. 2(b)).  When used with single-mode fibers, forked holograms have the dual property of both converting different orbital angular momentum beams into Gaussian beams travelling in different directions and filtering out all but the Gaussian part of the beam. Combined with polarizers and single-photon counting modules, the interferometers measure the single-photon Bell states (Fig. 2). Thus, by detecting the outputs of each of these “Spin-Orbit” controlled-NOT (CNOT) gates in coincidence, it was possible to perform a deterministic full polarization BSA.
\begin{figure}[btp]
\label{Spinorbit}
\subfigure{\includegraphics[width=0.95\linewidth]{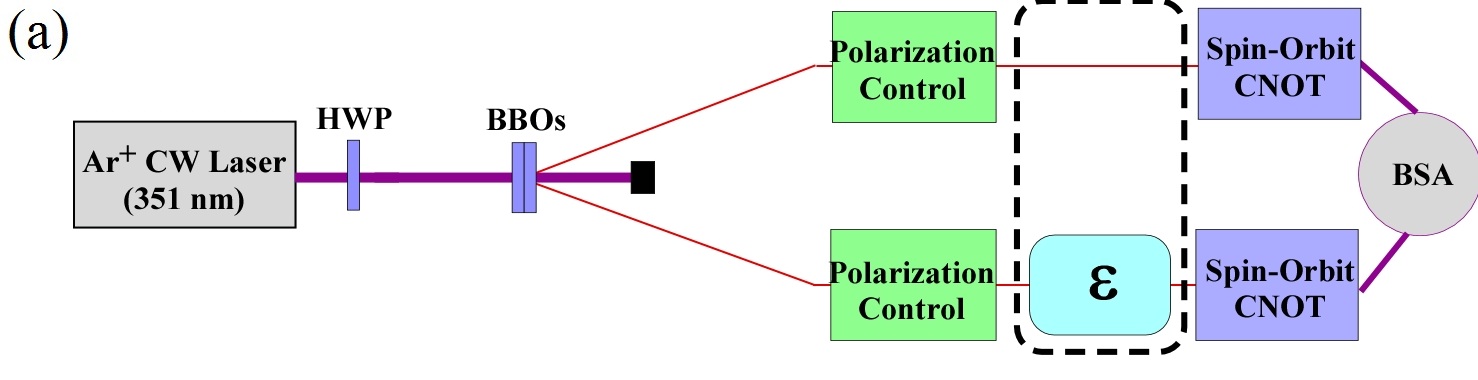}}\\
\subfigure{\includegraphics[width=0.95\linewidth]{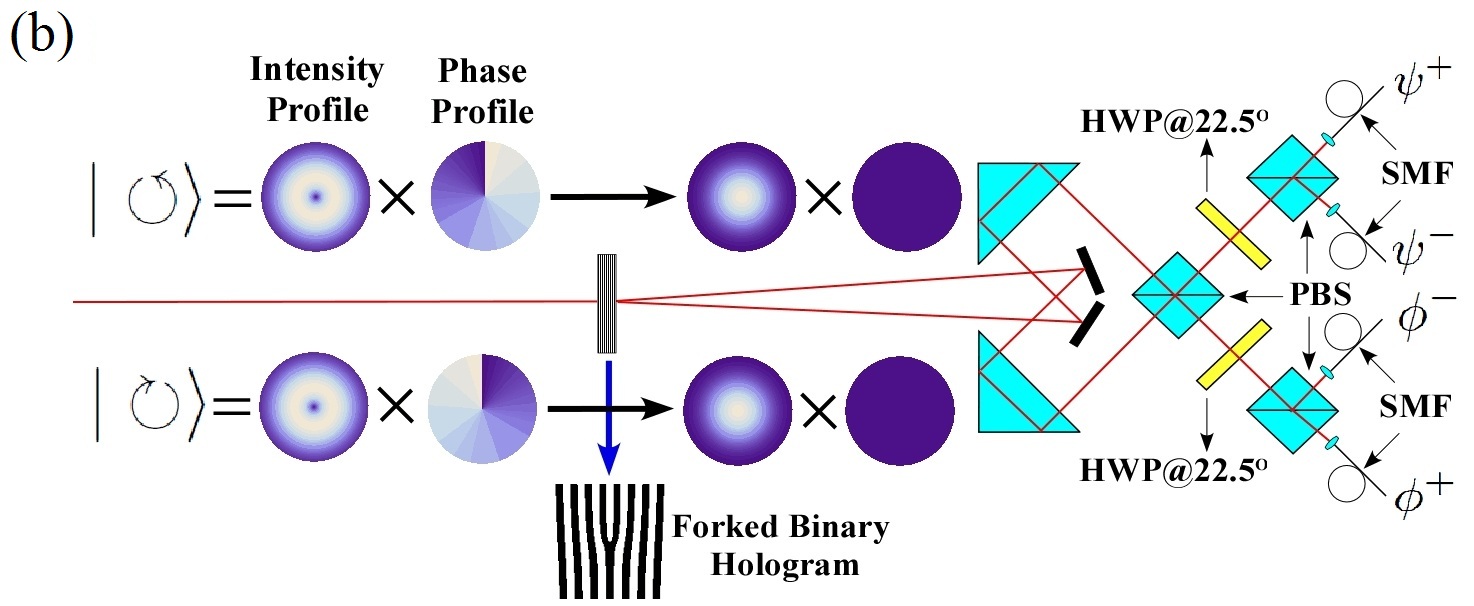}}
\caption{(a). Experimental setup used to perform DCQD on various single-photon quantum processes. A half-wave plate (HWP) tunes the pump polarization to generate photon pairs for SQPT or DCQD. The DCQD input states were prepared with liquid crystals (Polarization Control).  (b). The spin-orbit CNOT gates, which were used to measure the four single-photon hybrid-entangled Bell states in Eq. (5).  The outputs of a forked binary hologram are combined on a polarizing beamsplitter (PBS) and then spatially filtered with single-mode fiber (SMF).}
\end{figure}

We used DCQD to characterize several single-photon polarization processes, implemented using a variety of optical elements\footnote{One interesting advantage of using hyperentangled photon states is that it is possible to measure two-qubit quantum processes (e.g., which act on the polarizations of both the signal and idler photons) by making only minor modifications of the experimental setup in procedure.  This is due to the fact that DCQD requires only pair-wise BSA between the four qubits.  Just as the polarization Bell state (while keeping the orbital angular momentum in a particular Bell state) could be expressed as a superposition of single-photon Bell states, similar expressions may be made to perform BSA between all four combinations of qubits.}.  Pauli matrix rotation quantum processes were applied using half-wave plates, dephasing and depolarization processes were implemented using thick birefringent quartz plates \cite{decohere}, and the polarization quantum process was implemented with a sheet polarizer.  Tunable partial polarization/dephasing processes were implemented by placing polarization/dephasing processes in the beam path part of the count time and the identity process the remainder.

The $\chi$-matrices of each of the quantum processes and the identity were characterized using 18-experimental-configuration SQPT and 4-experimental-configuration DCQD.  Though in both techniques it is in principle possible to obtain the $\chi$-matrix from the recorded measurements through linear inversion methods, in neither case are the results constrained to the space of physically possible solutions.  For this reason maximum likelihood techniques \cite{maxlike} were used in both cases. The $\chi$-matrices calculated using both DCQD and SQPT were compared using Jamiolkowski fidelity $F_J$ \footnote{$F_J=Tr\left[\sqrt{\sqrt{[I\otimes\epsilon_1](\rho_\Phi)}[I\otimes\epsilon_2](\rho_\Phi)\sqrt{[I\otimes\epsilon_1](\rho_\Phi)}}\right]$, where quantum processes $\epsilon_1$ and $\epsilon_2$ act on one qubit of some two-qubit maximally entangled state, $\rho_\Phi$, and the fidelity is taken between the outputs of the two processes.}.  $F_J$ was chosen over other process fidelities because it is a stable metric and which can be calculated for all classes of quantum processes (in contrast to the more commonly used average process fidelity) \cite{fidelity}.

\begin{figure}[tp]
\label{data}
\subfigure{\includegraphics[width=0.32\linewidth]{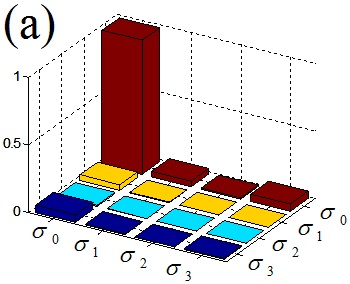}}
\subfigure{\includegraphics[width=0.32\linewidth]{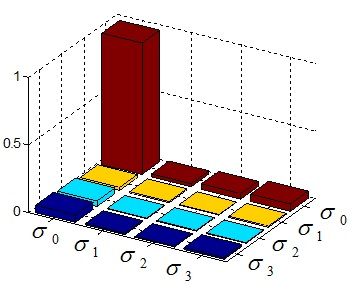}}
\begin{tabular}{c}$F_J=97.0 \pm 0.9\%$
\end{tabular}
\subfigure{\includegraphics[width=0.32\linewidth]{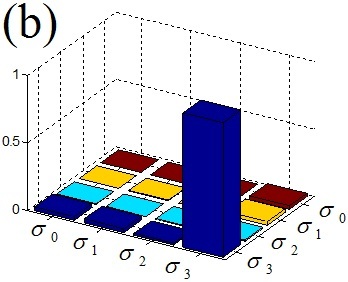}}
\subfigure{\includegraphics[width=0.32\linewidth]{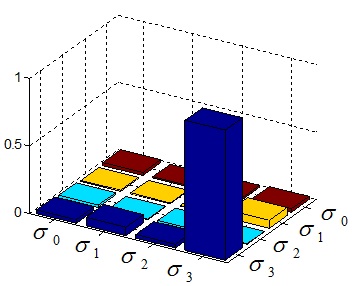}}
\begin{tabular}{c}$F_J=96.6 \pm 0.8\%$
\end{tabular} 
\subfigure{\includegraphics[width=0.32\linewidth]{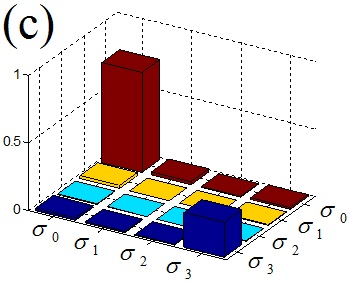}}
\subfigure{\includegraphics[width=0.32\linewidth]{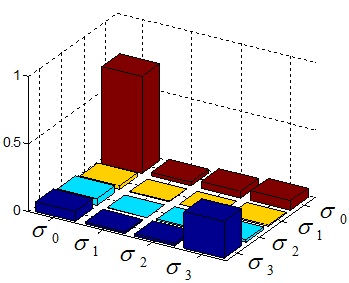}}
\begin{tabular}{c}$F_J=98.2 \pm 0.5\%$
\end{tabular}
\subfigure{\includegraphics[width=0.32\linewidth]{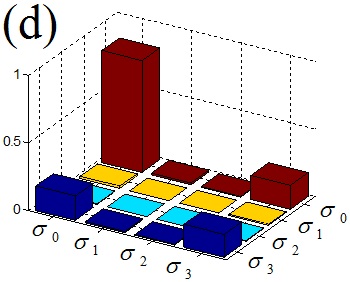}}
\subfigure{\includegraphics[width=0.32\linewidth]{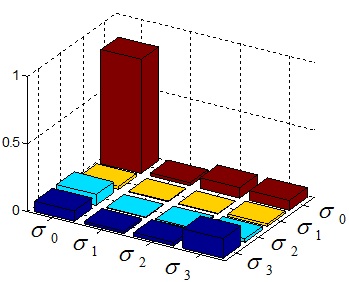}}
\begin{tabular}{c}$F_J=97.3 \pm 0.5\%$
\end{tabular}
\subfigure{\includegraphics[width=0.32\linewidth]{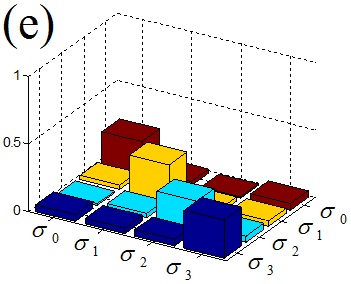}}
\subfigure{\includegraphics[width=0.32\linewidth]{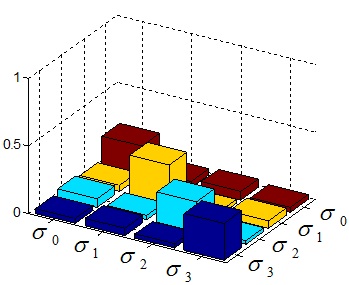}}
\begin{tabular}{c}$F_J=96.4 \pm 1.0\%$
\end{tabular}
\subfigure{\includegraphics[width=0.32\linewidth]{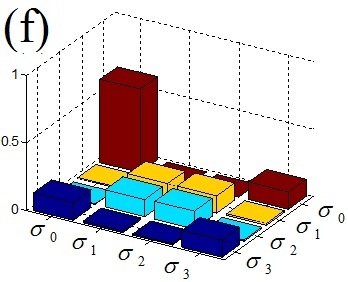}}
\subfigure{\includegraphics[width=0.32\linewidth]{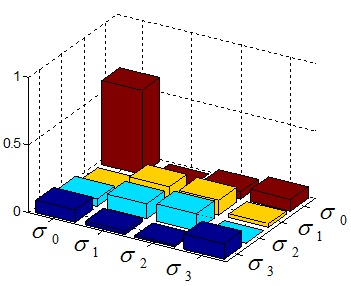}}
\begin{tabular}{c}$F_J=97.5 \pm 0.9\%$
\end{tabular}
\caption{Reconstructed $\chi$-matrices (absolute values of elements) from SQPT (left column) and DCQD (right column) measurements for the following: identity (a), $\sigma_z$ rotation (b),  partial dephasing (c)),  partial polarizer (d), depolarization (e), and simultaneous spin-lattice and spin-spin relaxation (f).  Each of the SQPT process matrices were characterized with 18 experimental settings while the DCQD process matrices were characterized with only 4.  The fidelities achieved with the latter after compensation of errors in measurement are also shown.}
\end{figure}

Initial $\chi$-matrices from the two methods showed less than desired agreement due to errors in the BSA.  However, it has been shown that errors introduced by imperfect DCQD state preparation and faulty BSA can be decoupled from the quantum process data if both error maps are well characterized \cite{GDCQD}, but to characterize the relevant two-qubit error map requires, a two-qubit QPT (which in general requires quadratically more experimental configurations than a single-qubit QPT).  Fortunately, because single-qubit DCQD only requires four experimental configurations, most of the information in these error maps is not required.  In fact, it is possible to completely compensate systematic errors by characterizing the input and measurement states with quantum state tomography (45 \footnote{(4 input states)$\times$(9 experimental configurations) + (1 measurement setting)$\times$(9 experimental configurations) = 45 experimental configurations} experimental configurations instead of 288 \footnote{(2 two-qubit quantum processes)$\times$($12^2$ experimental configurations) = 288 experimental configurations}).  This comparatively small number of experimental settings required to characterize systematic error is an advantage compared to AAPT techniques, which require between 54 and 288 experimental configurations, depending on measurement strategy (but a disadvantage compared with SQPT which requires only 24 experimental configurations). After correcting the measured systematic errors in the BSA, the overlap between identical processes when measured with SQPT and DCQD in general was significantly improved, from a mean $F_J$ of 89.4\% without correction to 96.1\% with it (Fig. 3).  The remaining errors appear to be the result of time-dependent drifts, which cannot be compensated for using the techniques described here.

It is possible to use DCQD techniques to measure some significant process parameters without performing a full process characterization.  For example, it is possible to measure both spin-lattice ($T_1$) and spin-spin ($T_2$) relaxation times using a single experimental setting \cite{DCQD, reduce}.  Though vitally important in characterizing the coherence of many types of quantum systems (atoms, ions, quantum dots, superconducting qubits, etc.), spin-lattice relaxation is not present in most photonic systems because photons do not interact strongly with thermodynamic reservoirs or decay to lower energy states.  However, we were able to simulate these processes by employing a combination of time-varying polarization processes.  This was accomplished by transmitting the photons through a thick quartz plate and a polarizer part of the time and through empty space (the identity process) the remainder.  While the individual values of $T_1$ and $T_2$ have little physical significance in our optical simulation, the ratio of $R \equiv T_2:T_1$ is a measure of how coherently states evolve in this process.  This ratio was measured with SQPT ($1.01 \pm 0.03$) using an over-complete set of 18 experimental settings and with DCQD using a \textit{single} experimental setting ($0.99 \pm 0.06$).  This process was also characterized with DCQD using all four experimental configurations (Fig. 3f).

In summary, we experimentally implemented a powerful technique for reducing the number of experimental configurations required for characterizing quantum processes through the use of hyperentanglement-assisted BSA.  Furthermore, we demonstrated how these processes could be characterized even when the measurement system is subject to systematic error.  Direct quantum process tomography approaches as presented here could be of significance for selectively estimating important biological or chemical properties in biomolecular complexes and nano-scale energy transfer systems \cite{bio} and monitoring non-Markovianity of quantum processes \cite{markov, markov2}.  In general, it should be possible to further decrease the number of experimental configurations required to characterize certain classes of quantum processes, by combining DCQD techniques with ``compressed sensing'' methods that have already been used to characterize sparse quantum states and processes with far fewer experimental configurations than would normally be required \cite{compress, compress2}.  We have also begun studying the number of state copies DCQD needs (compared to other QPT techniques) to statistically constrain the error of process estimation.  Our current numerical simulations indicate that DCQD requires fewer state copies that SQPT to constrain the error some classes of unitary processes; however, further investigation is required to quantify in general which QPT technique requires fewer state copies to constrain the error of different classes of quantum processes.

This work was funded by NSF Grant No. PHY-0903865, the ADNA/S\&T-IARPA project Hyperentanglement-Enhanced Advanced Quantum Communication (NBCHC070006), QuISM MURI Program, and DARPA QuBE program.

\vspace{23 mm}
\end{document}